\documentclass[twocolumn,showpacs,preprintnumbers,amsmath,amssymb]{revtex4}

\usepackage{graphicx}
\usepackage{amssymb}
\usepackage{amsmath}
\usepackage{times}
\usepackage{tabularx}
\usepackage{txfonts}
\usepackage{lineno}
\usepackage{multirow,bigdelim}
\usepackage{textcomp}
\usepackage{comment}
\usepackage{lineno}
\usepackage{float}
\usepackage[caption=false]{subfig}
\usepackage{subfig}
\usepackage{amsmath}
\usepackage{amstext}
\usepackage{graphicx}
\usepackage{tabularx,ragged2e}

 \usepackage{ifpdf}
 \ifpdf
 \usepackage[pdftex]{hyperref}
 \else
 \usepackage[hypertex]{hyperref}
 \fi

 \hypersetup{
   pdftitle={},
   pdfauthor={},
   pdfsubject={},
   pdfkeywords={},
   pdfstartview={},
   bookmarksopen=true, breaklinks=true, debug=true,
   colorlinks=true, linkcolor=red, citecolor=blue, urlcolor=blue
   }

\newcount\savefnused
\newcount\savefndone

\newcommand{\savefootnote}[2][\empty]
{\ifx\empty#1\footnotemark\else\footnotemark[#1]\fi
 \global\advance\savefnused by 1
 \expandafter\xdef\csname savefnmark\the\savefnused\endcsname{\thefootnote}
 \expandafter\xdef\csname savefntext\the\savefnused\endcsname{#2}
}
\newcommand{\flushfootnote}{\loop\ifnum\savefndone<\savefnused
  \global\advance\savefndone by 1
  \footnotetext[\csname savefnmark\the\savefndone\endcsname]
    {\csname savefntext\the\savefndone\endcsname}%
  \global\expandafter\let\csname savefnmark\the\savefndone\endcsname\relax
  \global\expandafter\let\csname savefntext\the\savefndone\endcsname\relax
\repeat}

\newcolumntype{Y}{>{\centering\arraybackslash}X}

\def\Slash#1{\ooalign{\hfil/\hfil\crcr$#1$}}

\begin{document}

\preprint{}

\title{Effect of the Pauli exclusion principle in the electric dipole moment of $^9$Be with $|\Delta S|=1$ interactions}

\author{Jehee Lee$^{1,2}$}
\author{Nodoka Yamanaka$^{3,2}$}
\author{Emiko Hiyama$^{4,2}$}
\affiliation{$^1$Department of Physics, Tokyo Institute of Technology, Meguro, Tokyo 152-8551, Japan}
\affiliation{$^2$Nishina Center, RIKEN, Saitama 351-0198, Japan}
\affiliation{$^3$IPNO, CNRS-IN2P3, Univ. Paris-Sud, Universit\'e Paris-Saclay,
91406 Orsay Cedex, France}
\affiliation{$^4$Department of Physics, Kyushu University, Fukuoka, 819-0395, Japan}

\begin{abstract}
We calculate the contribution of the $|\Delta S|=1$ $K$ meson exchange process generated by the Cabibbo-Kobayashi-Maskawa matrix to the electric-dipole moment (EDM) of the $^9$Be nucleus by considering the $\alpha n$-$\alpha \Lambda$ channel coupling.
It is found that the effect of the Pauli exclusion principle is not important intermediate $S=-1$ state, and that the result is consistent with the EDM of $^9$Be calculated with the $|\Delta S|=1$ interactions as a perturbation without considering the nucleus-hypernucleus mixing.
Our result suggests that the effect of the $|\Delta S|=1$ interactions is neither suppressed nor  enhanced in nuclei, if the difference of binding energies between the nucleus and the hypernucleus is small compared with the hyperon-nucleon mass difference.
\end{abstract}

\date{\today}

\maketitle

\section{\label{sec:intro}Introduction}

The electric-dipole moment (EDM) \cite{He:1990qa,Bernreuther:1990jx,Barr:1992dq,Jungmann:2013sga,NaviliatCuncic:2012zza,Khriplovich:1997ga,Ginges:2003qt,Pospelov:2005pr,Raidal:2008jk,Fukuyama:2012np,Engel:2013lsa,Yamanaka:2014,Roberts:2014bka,deVries:2015gea,Yamanaka:2016umw,Yamanaka:2017mef,Chupp:2017rkp,Safronova:2017xyt,Orzel:2018cui}
is often quoted as the most sensitive observable to the CP violation beyond standard model (SM) which is required to explain the baryon number asymmetry of the universe \cite{Sakharov:1967dj,Farrar:1993hn,Huet:1994jb}, and active searches using various systems such as the neutron \cite{Baker:2006ts}, atoms \cite{rosenberry,Regan:2002ta,Bishof:2016uqx,Graner:2016ses}, molecules \cite{Hudson:2011zz,Kara:2012ay,Baron:2013eja, Cairncross:2017fip, Andreev:2018ayy}, or muons \cite{Bennett:2008dy}, are currently being carried out.
There are also new ideas to measure it in paramagnetic atoms by using three-dimensional optical lattices \cite{Chin:2001zz,Sakemi:2011zz}, protons and light nuclei using storage rings \cite{Yamanaka:2016umw,Khriplovich:1998zq,Farley:2003wt,Orlov:2006su,Anastassopoulos:2015ura,Hempelmann:2017zgg}, strange and charmed baryons using bent crystals \cite{Botella:2016ksl,Baryshevsky:2018dqq}, $\tau$ leptons from the precision analysis of collider experimental data \cite{Chen:2018cxt,Koksal:2018env, Koksal:2018xyi}, electrons using polar molecules and inert gas matrices \cite{Vutha:2017pej}, etc.
It is also a probe of the axions \cite{Stadnik:2017hpa, Dzuba:2018anu}, which were first conceived to resolve the strong charge-parity (CP) problem \cite{Peccei:1977hh} and are now extensively discussed in the context of dark matter, or the Lorentz violation \cite{Araujo:2018lyw}.

One of the most attractive advantages of the EDM is that the effect of the CP phase of the Cabibbo-Kobayashi-Maskawa (CKM) matrix \cite{Kobayashi:1973fv}, which is the representative CP violation of the SM, is extremely small, at least for all known systems.
The CKM contributions to the EDM of light quarks \cite{Shabalin:1978rs,Shabalin:1980tf,Khriplovich:1985jr,Czarnecki:1997bu} and charged leptons \cite{Pospelov:1991zt,Booth:1993af,Pospelov:2013sca} appear from the three- and four-loop levels, respectively, due to the antisymmetry of the Jarlskog invariant \cite{Glashow:1970gm,Jarlskog:1985ht}, and thus are explicitly shown to be very small.
The Weinberg operator is also very small, with an estimated effect to the neutron EDM to be of $O(10^{-40})e$ cm \cite{Pospelov:1994uf}.

The CKM contributions to the EDM of composite systems are believed to be more enhanced due to the {\it long distance effect} \cite{Khriplovich:1981ca,McKellar:1987tf,Seng:2014lea,Pospelov:2013sca,Yamanaka:2015ncb}, where the Jarlskog combination is realized with two distinct $|\Delta S|=1$ hadron level interactions.
As for the nucleon EDM, this contribution is larger than the quark EDM contribution by two or three orders of magnitude, but the hadronic and nuclear level uncertainties are also large.
An important systematics of nuclear systems is the mixing of the $S=0$ nuclear state with the $S=-1$ hypernucleus through the weak interaction.
This effect has recently been evaluated for the deuteron \cite{Yamanaka:2016fjj}, and it was found to not be enhanced.

The story might however change for heavier nuclei, since the structure of hypernuclei significantly differs from the $S=0$ structure due to the relevance of the Pauli exclusion principle \cite{Hiyama:1997ub,Hiyama:2000jd,Hiyama:2009zz,Hiyama:2012gx,Hiyama:1996gv, Hiyama:1999me, Hiyama:2001xx, Hiyama:2002yj, Hiyama:2009ki, Hiyama:2010zzd, Hiyama:2010zzb, Hiyama:2010zz, Hiyama:2015bta}.
It has actually been shown in the study of $^{13}$C that the nuclear EDM is very sensitive to the change of the nuclear structure in parity transition \cite{Yamada:2015jha,Yamanaka:2016itb}.
This aspect is roughly controlled by the energy difference among transitioning states and the overlap of the matrix elements of operators contributing to the EDM.
In our case, we are interested in the second one, since the energy difference is roughly given by the hyperon-nucleon mass splitting.
If the effect of the Pauli blocking is important, the transition matrix elements, and consequently the EDM, might be significantly suppressed.

The purpose of this paper is to test whether the Pauli exclusion principle affects the nuclear EDM generated by the $|\Delta S|=1$ interactions through the nucleus-hypernucleus mixing.
For that, we choose the $^9$Be nucleus whose structure, together with that of $^9_\Lambda$Be, is well known from the cluster model \cite{Hiyama:1997ub,Hiyama:2000jd,Funaki:2015uya,Yamanaka:2016umw,Yamanaka:2015qfa}.
This nucleus is of particular interest because it is on the border of shell-like and cluster structures. The $^9$Be nucleus is then a prototype of all other heavier stable odd nuclei which have core plus valence configurations.
The result of our work then also has an impact in the estimation of the EDM of other interesting systems such as heavy atoms and nuclei \cite{Khriplovich:1997ga,Ginges:2003qt,Engel:2013lsa,Roberts:2014bka,Yamanaka:2016umw,Yamanaka:2017mef,Chupp:2017rkp,Safronova:2017xyt}, or in the analysis of $T$-odd angular correlations of nuclear beta decay \cite{Herczeg:1997se,Herczeg:2001vk,Gonzalez-Alonso:2018omy,Nair:2018mwa}.
It is also important to note that this influences the sensitivity of the above observables on general $|\Delta S| = 1$ processes, important in the phenomenological analysis of new physics beyond the standard model with flavor violation \cite{Ellis:1981ts,
Cheng:1987rs,
Hagelin:1992tc,
Gabbiani:1996hi,
Buchalla:1996fp,
Fritzsch:1999ee,
DAmbrosio:2002vsn,
Buras:2003td,
Agashe:2004cp,
Antonelli:2009ws,
Isidori:2010kg,
Buras:2010zm,Charles:2015gya,
Kitahara:2016nld,Kitahara:2016otd,Cirigliano:2016yhc,Endo:2017ums,Smith:2017dtz,Chang:2017wpl,Agrawal:2017evu,Aebischer:2018csl,Dekens:2018bci,
Gisbert:2017vvj, Haba:2018byj, Chen:2018ytc, Chen:2018vog, Bailey:2018feb,Haba:2018rzf}.

This paper is organized as follows.
In the next section, we describe the quark level $|\Delta S| = 1$ weak effective Hamiltonian.
In Sec. \ref{sec:interaction}, we present the setup of the $N N$ and $N \Lambda$ interactions and the $\alpha$ cluster model used in this work.
In Sec. \ref{sec:power}, we estimate the EDM using power counting rules.
We then explain the Gaussian expansion method (GEM) which is used to calculate the nuclear structure and the formulation of the EDM in Secs. \ref{sec:GEM} and \ref{sec:EDM}, respectively.
The results are presented and discussed in Sec. \ref{sec:results}.
We summarize our paper in Sec. \ref{sec:summary}.

\section{Quark level weak effective hamiltonian}\label{sec:quark}

\begin{figure}[]
\begin{center}
\includegraphics[scale=0.5]{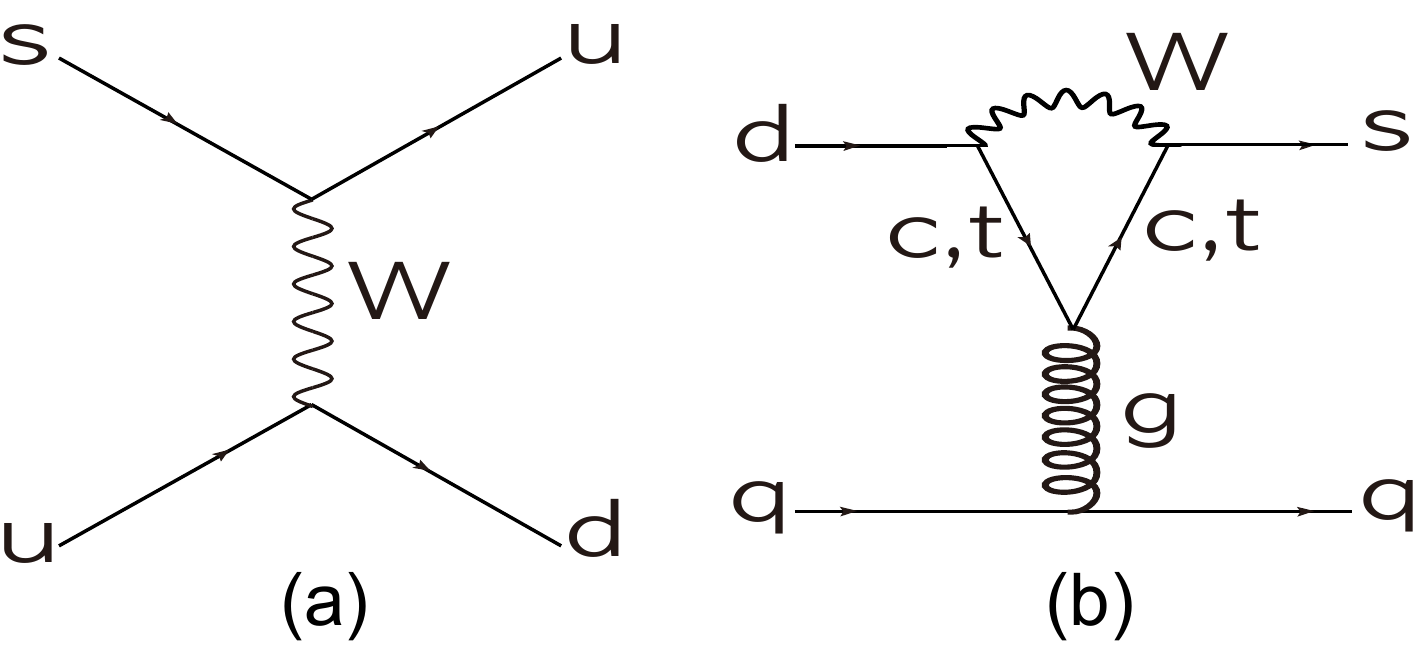}
\caption
{$| \Delta S |=1$ $W$ boson exchange processes, with
(a) the tree level diagram, and
(b) the penguin diagram.
}\label{fig:w_ex}
\end{center}
\end{figure}

In the standard model, the leading CP violation is generated by two $W$ boson exchanges for which the couplings with quarks fulfill the Jarlskog combination \cite{Jarlskog:1985ht}.
As seen in the Introduction,
the long distance contribution is dominant with the EDM.
We therefore need two distinct $| \Delta S |=1$ four-quarks
interactions.
For example, the $| \Delta S |=1$ $W$ boson exchange processes are shown in Fig. \ref{fig:w_ex}.

After the integration of the $W$ boson,
the $|\Delta 􏱖S| = 1$ effective Hamiltonian of the SM is given by
\begin{align}\label{}
& \mathcal{H}_{eff}(\mu=m_{W}) \nonumber \\
&= \frac{G_{F}}{\sqrt{2}}
\Bigg\{ C_1 (\mu=m_{W}) \left[ V^*_{us}V_{ud}Q^u_1
+V^*_{cs}V_{cd}Q^c_1\right] \nonumber \\
& + C_2(\mu=m_{W}) \left[ V^*_{us}V_{ud}Q^u_2
+V^*_{cs}V_{cd}Q^c_2\right] \nonumber \\
&
-V^*_{ts}V_{td}\sum^{6}_{i=3} C_i (\mu=m_{W}) Q_i
\Bigg\} + \rm{(h.c.)},
\end{align}
where $V_{qq'}$ are the CKM matrix elements
and the Fermi constant is
$G_{F}=1.16637 \times 10^{-5} \; \rm{GeV}^{-2}$
\cite{Tanabashi:2018oca}.
The mass of the $W$ boson is $m_W=80.4$ GeV \cite{Tanabashi:2018oca}.
Moreover, the $| \Delta S |=1$ four-quark operators $Q_{i}$ with $i=1 \sim 6$
are defined in the following basis \cite{Buras:1991jm,Buchalla:1995vs}
\begin{align}\label{}
Q_{1}^{q}
&=\bar{s}_{\alpha} \gamma^{\mu} (1 - \gamma_5)q_{\beta} \cdot
\bar{q}_{\beta} \gamma_{\mu} (1 - \gamma_5) d_{\alpha} , \\
Q_{2}^{q}
&=\bar{s}_{\alpha} \gamma^{\mu} (1 - \gamma_5)q_{\alpha} \cdot
\bar{q}_{\beta} \gamma_{\mu} (1 - \gamma_5) d_{\beta} , \\
Q_{3}^{q}
&=\bar{s}_{\alpha} \gamma^{\mu} (1 - \gamma_5)d_{\alpha} \cdot
\sum^{N_{f}}_{q} \bar{q}_{\beta} \gamma_{\mu}
(1 - \gamma_5) q_{\beta}, \\
Q_{4}^{q}
&=\bar{s}_{\alpha} \gamma^{\mu} (1 - \gamma_5)d_{\beta} \cdot
\sum^{N_{f}}_{q} \bar{q}_{\beta} \gamma_{\mu}
(1 - \gamma_5) q_{\alpha}, \\
Q_{5}^{q}
&=\bar{s}_{\alpha} \gamma^{\mu} (1 - \gamma_5)d_{\alpha} \cdot
\sum^{N_{f}}_{q} \bar{q}_{\beta} \gamma_{\mu}
(1+ \gamma_5) q_{\beta}, \\
Q_{6}^{q}
&=\bar{s}_{\alpha} \gamma^{\mu} (1 - \gamma_5)d_{\beta} \cdot
\sum^{N_{f}}_{q} \bar{q}_{\beta} \gamma_{\mu}
(1 +\gamma_5) q_{\alpha},
\end{align}
where $\alpha$ and $\beta$ denote the color indices of the quarks.
The Wilson coefficients $C_i$ are evolved down to the hadronic scale
according to the next-to-next leading logarithmic approximation
of the renormalization-group equation
 \cite{Buras:1991jm,Buchalla:1995vs,Yamanaka:2016itb}.
 The effective Hamiltonian near the hadronic scale $\mu=1$ GeV
 is given by
 \begin{align}\label{}
 \mathcal{H}_{eff} (\mu)
 &= \frac{G_F}{\sqrt{2}} V^{*}_{us} V_{ud}
 \sum^{6}_{i=1} [z_i(\mu) + \tau y_i(\mu)] Q_i(\mu)
 + \rm{(h.c.)},
 \end{align}
 with $\tau \equiv - \frac{V^{*}_{ts} V_{td}}{V^{*}_{us} V_{ud}}$.
 After the renormalization down to the hadronic scale $\mu = 1$ GeV,
 the Wilson coefficients $y_i$ and $z_i$ ($i=1\sim6$) are determined as
\begin{align}\label{}
z(\mu=1\;\rm{GeV}) &=
\left(
\begin{array}{c}
-0.107 \\
1.02 \\
1.76 \times 10^{-5} \\
-1.39 \times 10^{-2} \\
6.37 \times 10^{-3} \\
-3.45 \times 10^{-3}
\end{array}
\right), \\
y(\mu=1\;\rm{GeV}) &=
\left(
\begin{array}{c}
0\\
0 \\
1.48 \times 10^{-2} \\
-4.81 \times 10^{-2} \\
3.22 \times 10^{-3} \\
-5.69 \times 10^{-2}
\end{array}
\right). \label{eq:y}
\end{align}
\begin{figure}[]
\begin{center}
\includegraphics[scale=0.6]{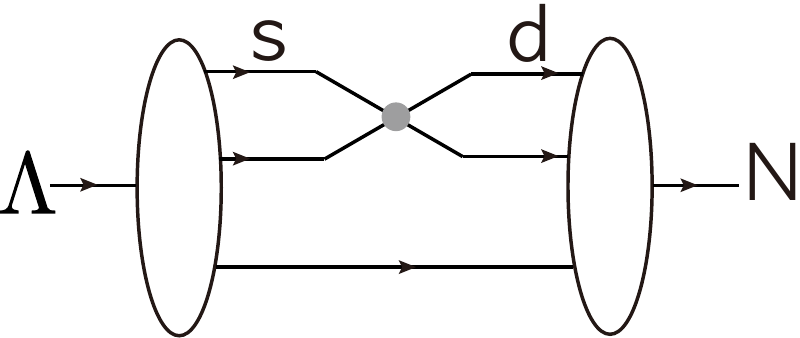}
\caption{
Schematic pictures of the $\Lambda$-$N$ transition from the $|\Delta S|=1 $
four-quark interactions.
}\label{fig:trans}
\end{center}
\end{figure}
From the above Wilson coefficients, we formulate the $|\Delta S|=1$ interactions.

Let us first derive the hyperon-neutron transition.
The effective Hamiltonian of this one-body process is expressed by
\begin{align}\label{}
T^{(|\Delta S|=1)}
&= -a_{n \Lambda} [n^{\dagger} \Lambda] +(\rm{h.c.}),
\end{align}
where $a_{n \Lambda}$ is the weak-coupling constant
of the $\Lambda$-nucleon transition, which is given in terms of the hadron matrix element as
\begin{align}\label{}
a_{n \Lambda}
&= | V_{us} V_{ud} |  \frac{G_F}{\sqrt{2}}
	(z_1 - z_2) \langle n | Q_{2}^{NR} | \Lambda \rangle .
\end{align}
Here the baryon scalar density matrix is given by
\begin{align}\label{}
\langle n | \bar{d} s  | \Lambda \rangle
\approx \sqrt{\frac{3}{2}} \frac{m_N - m_{\Lambda}}{m_s}
\approx -1.80,
\end{align}
where the renormalization scale is $\mu=1$ GeV
and the strange quark mass, the nucleon mass, and the Lambda mass are
$m_s=120$ MeV, $m_N=938$ MeV and $m_{\Lambda}=1115.6$ MeV, respectively \cite{Tanabashi:2018oca}.
Here we use the $\Lambda$-neutron transition matrix element calculated in Ref. \cite{Hiyama:2004wd}
\begin{align}\label{}
\langle n | Q_{2}^{NR} | \Lambda \rangle
&= -9.65 \times 10^{-3} \; \rm{GeV}^{3}.
\end{align}
where $Q_{2}^{NR}$ is the nonrelativistic reduction of $Q_2^q$.
This result was obtained by calculating the nonleptonic hyperon decay with $Q_{2}^{NR}$as input in the quark model (see Fig. \ref{fig:trans}).

\begin{figure}[]
\begin{center}
\includegraphics[scale=0.7]{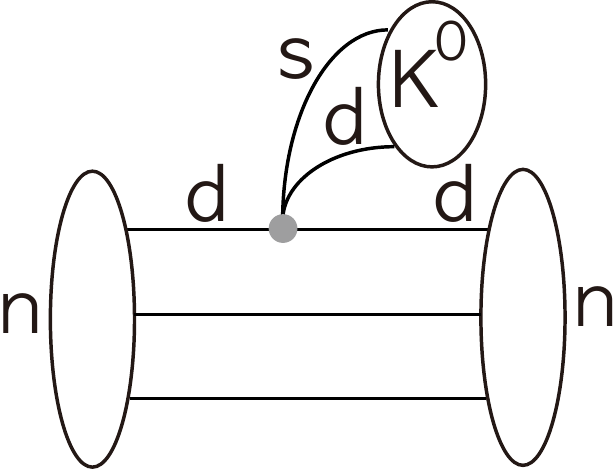}
\caption{
Schematic picture of the contribution of the $|\Delta S|=1$ four-quark interactions to $|\Delta S|=1$ meson-baryon interactions
using the factorization.
}\label{fig:Npi}
\end{center}
\end{figure}
We now derive the $|\Delta S|=1$ meson-baryon interaction.
By using the factorization approach (see Fig. \ref{fig:Npi}),
we obtain the $|\Delta S|=1$ P-odd kaon-nucleon interaction
\begin{align}\label{}
\mathcal{L}_{K^0 NN}
&= \bar{g}_{K^0 pp} K^0 \bar{p}p
+ \bar{g}_{K^0 nn} K^0 \bar{n}n
+ \rm{(h.c.)},
\end{align}
where the couplings $\bar{g}_{K^0pp}$ and
$\bar{g}_{K^0nn}$ are expressed as
\begin{align}\label{}
\bar{g}_{K^0pp} & \approx
G_{y} \langle \bar{K}^0 | \bar{s} \gamma_5 d | 0 \rangle
\langle p | \bar{d} d| p \rangle, \\
\bar{g}_{K^0nn} & \approx
G_{y} \langle \bar{K}^0 | \bar{s} \gamma_5 d | 0 \rangle
\langle n | \bar{d} d| n \rangle,
\end{align}
with $G_{y} \equiv \frac{iJ}{|V_{ud}V_{us}|}
\left[ \frac{2}{3}y_5 +2y_6 \right] $
and the Jarlskog invariant
$J=(3.06^{+0.21}_{-0.20})\times 10^{-5}$
\cite{Jarlskog:1985ht,Tanabashi:2018oca}
while $y_5$ and $y_6$ are given in Eq. (\ref{eq:y}).
A more systematic chiral Lagrangian can be found in Refs.
\cite{He:1992jh,Feijoo:2018den}
The pseudoscalar matrix element can be transformed by using the partially conserved axial current formula as
\begin{align}\label{}
\langle \bar{K}^0 | \bar{s} \gamma_5 d | 0 \rangle
& \approx \frac{i}{\sqrt{2}f_{K}}
\langle 0 | \bar{q}q + \bar{s}s | 0 \rangle,
\end{align}
where $f_{K} = 1.2 f_\pi$ with the pion-decay constant $f_{\pi}=93$ MeV.
Following the Gell-Mann-Oakes-Renner relation, the chiral condensate is given by
\begin{align}\label{}
\langle 0 | \bar{q}q  | 0 \rangle
&= - \frac{m_{\pi}^2 f_{\pi}^2}{m_u+ m_d}=-(265\;\rm{MeV})^3,
\end{align}
where $m_u=2.2$ MeV and $m_d=4.7$ MeV
at the renormalization scale $\mu=2$ GeV \cite{Tanabashi:2018oca, Dominguez:2018azt, Campos:2018ahf,Bazavov:2018omf}.
Here the pion mass is $m_{\pi}=139$ MeV.
The chiral condensate of the strange quark is close to that of light quarks:
$\langle 0 |\bar{s}s | 0 \rangle \approx \langle 0 | \bar{q}q  | 0 \rangle$ \cite{McNeile:2012xh}.
The nucleon scalar matrix elements are given by
\begin{align}\label{}
  \langle p | \bar{d} d| p \rangle
+  \langle n | \bar{d} d| n \rangle
 & \approx 10,
 \end{align}
which is derived from $\sigma_{\pi N} \equiv \frac{1}{2} (m_u+m_d)
  \langle  N | \bar{u} u + \bar{d} d| N  \rangle
\approx 45 $ MeV.
Note that $\sigma_{\pi N}$ obtained from phenomenological extractions ($\simeq$ 60 MeV) \cite{Alarcon:2011zs,Hoferichter:2015dsa, Yao:2016vbz,RuizdeElvira:2017stg} and that from lattice calculations ($\simeq$ 30 MeV) \cite{Alexandrou:2017qyt,Yang:2015uis, Durr:2015dna, Bali:2016lvx, Yamanaka:2018uud} are not consistent, so we just took the average.

\section{Interaction}\label{sec:interaction}
The Hamiltonian of  $^9$Be and  $^9_{\Lambda}$Be
is given by
\begin{align}\label{}
H &= \sum^9_{a=1}T_a + \Delta M + V_{NN} + V_{YN}
+ V_{\mathrm{Pauli}}
\nonumber \\ & \;\;\;
+ \sum^9_{a=1} T_a^{(|\Delta S|=1)}
+ \mathcal{H}_{\Slash{P}}^{(|\Delta S|=1)},
\end{align}
with the kinetic energy $T$, the nuclear potential $V_{NN}$,
the hyperon-nucleon potential $V_{YN}$,
the strangeness violating weak one-body transition $T_a^{(|\Delta S|=1)}$,
and the $|\Delta S|=1$ P-odd meson exchange two-body potential
$\mathcal{H}_{\Slash{P}}^{(|\Delta S|=1)}$.
The mass shift $\Delta M = m_{\Lambda} -m_{N}$ is required to simultaneously consider the nucleus and the hypernucleus.

Let us first define the strangeness conserving sector.
We employ the $N$-$\alpha$ and $\alpha$-$\alpha$ interactions
which reproduce the scattering phase shift of
the $N$-$\alpha$ and $\alpha$-$\alpha$ systems at low energy
\cite{Hasegawa:1786,Kanada:1979}.
To reproduce the binding energy of $^9$Be (1.57 MeV), we introduced a small shift in the central $N$-$\alpha$ interaction.
For the $YN$ interaction, the YNG $\Lambda N$ interaction \cite{Yamamoto:1994tc} is employed.
It is parametrized as
\begin{align}\label{}
V_{\Lambda N} (r, k_F)
&= \sum_{i=1}^{3} \bigg[ (v_{0,even}^i + v_{\sigma \sigma, even}^i
\mbox{\boldmath $\sigma$}_{\Lambda}\cdot\mbox{\boldmath $\sigma$}_{N})\frac{1+P_r}{2}
\nonumber \\
& \;\;\;\; +(v_{0,odd}^i + v_{\sigma \sigma, odd}^i
\mbox{\boldmath $\sigma$}_{\Lambda}\cdot\mbox{\boldmath $\sigma$}_{N})\frac{1+P_r}{2}
\bigg] e^{-\left(\frac{r}{\beta_i}\right)^2},
\end{align}
where $P_r$ is the space exchange operator.
The strengths $v_{0,even}^i$, $v_{\sigma \sigma, even}^i$, $v_{0,odd}^i$, and $v_{\sigma \sigma, odd}^i$ are defined in Ref. \cite{Yamamoto:1994tc}.
By using this interaction, the energy of $^5_{\Lambda}$H is exactly reproduced,
$B(^5_{\Lambda}\mathrm{He})=3.12$ MeV.

The Pauli blocking between the $N$-$\alpha$ and $\alpha$-$\alpha$
systems is taken into account by the orthogonality condition model (OCM)
\cite{Saito:1968}.
The OCM projection operator $V_{\mathrm{Pauli}}$
is given by
\begin{align}\label{}
V_{\mathrm{Pauli}}
&=  \lim_{\lambda \to \infty} \lambda \sum_{f}
|\phi_f(r_{\alpha x}) \rangle \langle \phi_f(r_{\alpha x})|,
\end{align}
where $x=N$ or $\alpha$.
The operator rules out the amplitude of the forbidden
states in the $N$-$\alpha$ ($f=0s$) and $\alpha$-$\alpha$ ($f=0s, 1s, 0d$) systems
\cite{Kukulin:1995tsx}.
The Gaussian range parameter of the nucleon $0s$ orbit
in the $\alpha$-cluster is $b=1.358$ fm.

\begin{figure}[]
\begin{center}
\includegraphics[]{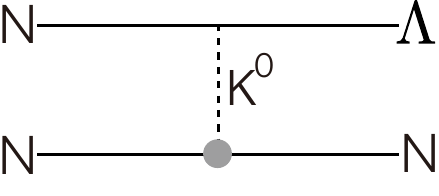}
\caption{
$|\Delta S|=1$ P-odd interbaryon force
by assuming the K$^0$ meson exchange.
}\label{fig:NL}
\end{center}
\end{figure}

\begin{figure}[]
\begin{center}
\includegraphics[scale=0.4]{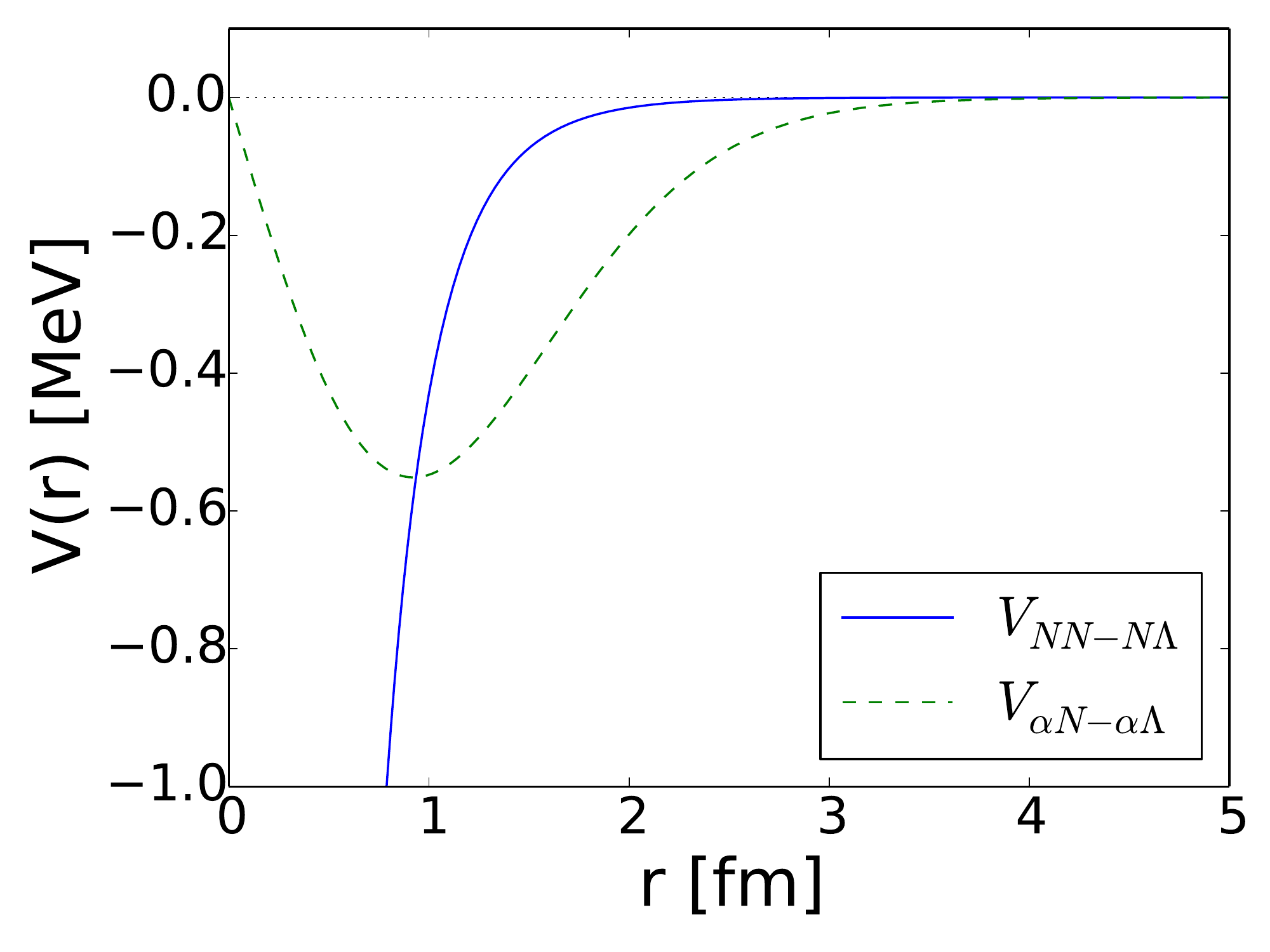}
\caption{
The radial shape of the folding $|\Delta S|=1$ kaon exchange $ NN$-$N  \Lambda$ and $\alpha N$-$\alpha N$ coupling potentials.}
\label{fig:potential}
\end{center}
\end{figure}

Now let us introduce the strangeness violating interactions.
We model the $|\Delta S|=1$ P-odd interbaryon force
by assuming the one-kaon exchange (see Fig. \ref{fig:NL}), which is the relevant one in this work.
The $|\Delta S|=1$ two-body interaction is given as
\begin{align}\label{}
\mathcal{H}_{\Slash{p}}^{|\Delta S|=1} &=
- g_{KN \Lambda}\bar{g}_{\bar{K}^0pp}
\left[ \Lambda^{\dagger} n \right]_2 \mbox{\boldmath $\sigma$}_2
\cdot \hat{\bf{r}} V_{pn-p \Lambda}(r)
\nonumber \\  & \;\;\;\;
-g_{KN \Lambda}\bar{g}_{\bar{K}^0nn}
\left[ \Lambda^{\dagger} n \right]_2 \mbox{\boldmath $\sigma$}_2
\cdot \hat{\bf{r}} V_{nn-n \Lambda}(r)
\nonumber \\ & \;\;\;\;
+(1 \leftrightarrow 2)
+(\mathrm{h.c.}),
\end{align}
where $\mbox{\boldmath $\sigma$}_2$ and $\left[ \Lambda^{\dagger} n \right]_2$ indicate the spin matrix and the strangeness transition operator of the second baryon, respectively, and $\hat{\bf{r}}$ is the unit vector directed from baryon 2 to baryon 1.
The P-even meson-baryon coupling is given by
$g_{K \Lambda N} = \frac{m_{N}+m_{\Lambda}}{2\sqrt{3} f_{\pi}}
(D+3F) \approx 13.6$, derived from
the leading terms of the chiral Lagrangian
\cite{Hisano:2004tf,Faessler:2006at,Fuyuto:2012yf,Yamanaka:2014}.
The coupling potential is defined by
\begin{align}\label{eq:potential}
V_{Nn-N \Lambda}(r) \hat{\bf{r}} &=
- \frac{1}{4 \mu_{N \Lambda}}
\frac{m_K}{4 \pi} \frac{e^{-m_K r}}{r}
\left( 1+ \frac{1}{m_K r}
\right) \hat{\bf{r}},
\end{align}
where $N=p$ or $n$ and the kaon mass is $m_K=497.6$ MeV \cite{Tanabashi:2018oca}.
The reduced mass is defined as
$\mu_{N \Lambda} \equiv \frac{m_N m_{\Lambda}}{m_N+m_{\Lambda}}$.
The nonlocal term in the $|\Delta S|=1$ $K$ meson exchange interaction is neglected since its effect is small ($ \sim O(10 \%)$).
The potential is displayed in Fig. \ref{fig:potential}.

\begin{figure}[]
\begin{center}
\includegraphics[width=\linewidth]{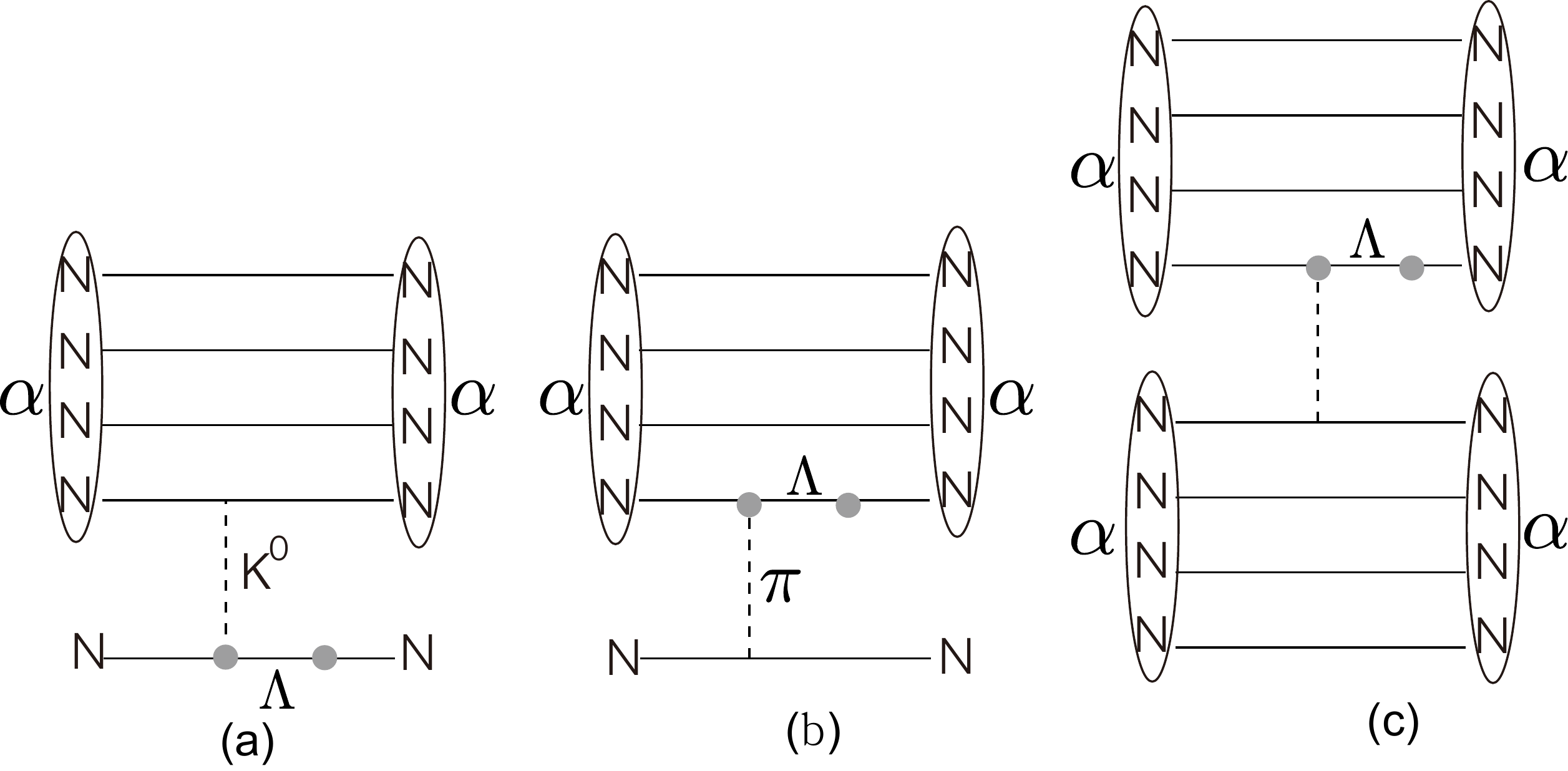}
\caption{From the left panel,
the $\alpha N$ interaction with the $K^0$ exchange,
the $\alpha N$ interaction with the $\pi$ exchange,
and the $\alpha \alpha$ interaction.
Only (a) survives whereas (b) and (c) are canceled
in the $\alpha$-cluster model.
}\label{Fig:aa&aN}
\end{center}
\end{figure}

In the $\alpha$-cluster model,
the relevant degrees of the freedom
are the baryon and the $\alpha$-cluster.
We therefore need to fold the $|\Delta S|=1$
two-body potential.
The folding procedure works as follows \cite{Yamanaka:2016umw}
\begin{align}\label{}
& V_{\alpha N - \alpha \Lambda} (r) {\bf{\hat{r}}} \nonumber \\
&= \frac{m_K}{ 2\sqrt{3} \pi^{\frac{3}{2}} b \mu_{N \Lambda}  }
\frac{{\bf{\hat{r}}}}{r}
\int^{\infty}_{0} dR'
e^{m_K R'} \left( 1 + \frac{1}{m_K R'}\right)
\nonumber \\ & \;\;\;\; \times
\left[ e^{-(\frac{r-R'}{b})^2}
\left( \frac{3 b^2}{8  rR'} -1\right)
-e^{-(\frac{r+R'}{b})^2}
\left( \frac{3 b^2}{8  rR'}+1\right)\right].
\end{align}
The radial shape of this potential is described in
Fig. \ref{fig:potential}.
It is important to note that the folding cancels
the $|\Delta S|=1$ two-body potential
in the case where the $\Lambda$ is created
by annihilating a nucleon in the $\alpha$-cluster
(see Fig. \ref{Fig:aa&aN}).
The $\eta$ and $\pi$ exchanges are not allowed due this and to the spin closure of the $\alpha$-cluster.
This is why only the $K^0$ exchange is possible in the $\alpha N$-$\alpha \Lambda$ channel coupling.
Moreover, since there is no spin and isospin in the $\alpha$ particle,
the $K^0$, $\eta$ and $\pi$ exchanges
are forbidden in the $\alpha \alpha$ interaction.

\section{Power counting estimate} \label{sec:power}
Let us estimate the EDM generated by the transition between $S=0,-1$ and opposite parity states by using power counting \cite{deVries:2011an}.
The EDM of $^9$Be in the leading-order perturbation is given by
\begin{widetext}
\begin{align}\label{}
d_{^9\mathrm{Be}} &=
\sum \hspace{-1.4em} \int_n \; \sum \hspace{-1.4em} \int_m \;
\frac{\langle\, ^9\mathrm{Be}(3/2^-) \,|\, \mathcal{H}_{\Slash{P}}^{(|\Delta S|=1)}\,|\,  ^9_{\Lambda}\mathrm{Be}^{(n)}(3/2^+) \,\rangle
\langle\, ^9_{\Lambda}\mathrm{Be}^{(n)}(3/2^+) \,|\, T_a^{(|\Delta S|=1)} \,|\,  ^9\mathrm{Be}^{(m)}(3/2^+) \,\rangle
\langle\, ^9\mathrm{Be}^{(m)}(3/2^+) \,|\, \mathcal{E} \,|\  ^9\mathrm{Be}(3/2^-) \,\rangle}
{\left(E[{^9\mathrm{Be}}(3/2^-)]-E[{^9_{\Lambda}\mathrm{Be}}^{(n)}(3/2^+)]\right)
 \left(E[{^9_{\Lambda}\mathrm{Be}}^{(n)}(3/2^+)]-E[{^9\mathrm{Be}}^{(m)}(3/2^+)]\right)} \nonumber \\
&+
\sum \hspace{-1.4em} \int_n \; \sum \hspace{-1.4em} \int_m \;
\frac{\langle\, ^9\mathrm{Be}(3/2^-) \,|\, \mathcal{H}_{\Slash{P}}^{(|\Delta S|=1)}\,|\,  ^9_{\Lambda}\mathrm{Be}^{(n)}(3/2^+) \,\rangle
\langle\, ^9_{\Lambda}\mathrm{Be}^{(n)}(3/2^+) \,|\, \mathcal{D} \,|\  ^9_{\Lambda}\mathrm{Be}^{(m)}(3/2^-) \,\rangle
\langle\, ^9_{\Lambda}\mathrm{Be}^{(m)}(3/2^-) \,|\, T_a^{(|\Delta S|=1)} \,|\,  ^9\mathrm{Be}(3/2^-) \,\rangle
}
{\left(E[{^9\mathrm{Be}}(3/2^-)]-E[{^9_{\Lambda}\mathrm{Be}}^{(n)}(3/2^+)]\right)
 \left(E[{^9_{\Lambda}\mathrm{Be}}^{(m)}(3/2^-)]-E[{^9\mathrm{Be}}(3/2^-)]\right)}
\nonumber \\
&+ (\mathrm{permutation}) ,
\label{eq:long}
\end{align}
\end{widetext}
where $\sum \hspace{-1.em} \int $ means that we take the sum for the bound states and the integral for the continuum states.
\begin{figure}[]
\begin{center}
\includegraphics[width=\linewidth]{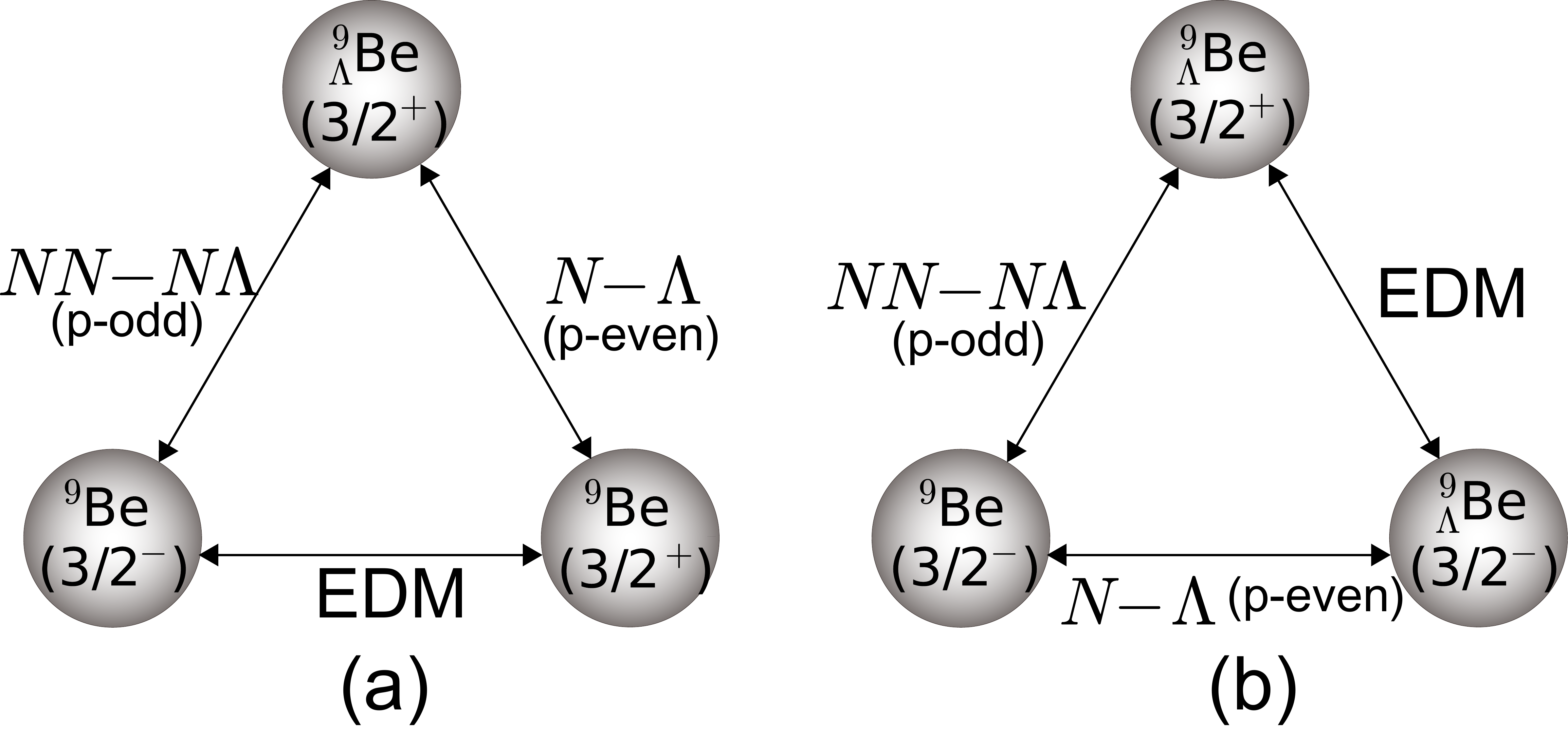}
\caption{
Schematic picture of transition between states contributing to the EDM of $^9$Be.
}\label{fig:1}
\end{center}
\end{figure}
$\mathcal{E}$ is the electric-dipole operator.
Let us first inspect the denominator. The one of the first term (see Fig. \ref{fig:1} (a)) is of $O(\Delta)$, with $\Delta\equiv m_\Lambda-m_N$, because $E[{^9\mathrm{Be}}^{(m)}(3/2^+)]$ can be close to $E[{^9_{\Lambda}\mathrm{Be}}^{(n)}(3/2^+)]$ due to the summation  over $m$ continuum states.
However, the second term (see Fig. \ref{fig:1} (b)) is of $O\left(\Delta^2\right)$ because $E[{^9\mathrm{Be}}^{(m)}(3/2^-)]$ is fixed.
This means that the first term of Eq. (\ref{eq:long}) is larger than the second term by a factor of $O (p^2/m_N \Delta)
\sim 1/20$ where $p$ is the soft scale of the order of typical binding momentum.
We also note that the nuclear EDM receives a contribution from the intrinsic EDM of the nucleon, which scales as $O(1/\Delta)$ in the chiral perturbation theory \cite{Seng:2014lea}.
From the power-counting argument, this can interfere with the first term of Eq. (\ref{eq:long}).
We will come back to this point later in the error estimation.
The other $|\Delta S|=1$ processes have less effect since the mass difference between strange and nonstrange hadrons is larger.

\section{Gaussian Expansion Method}\label{sec:GEM}
\begin{figure}[]
\begin{center}
\includegraphics[scale=0.45]{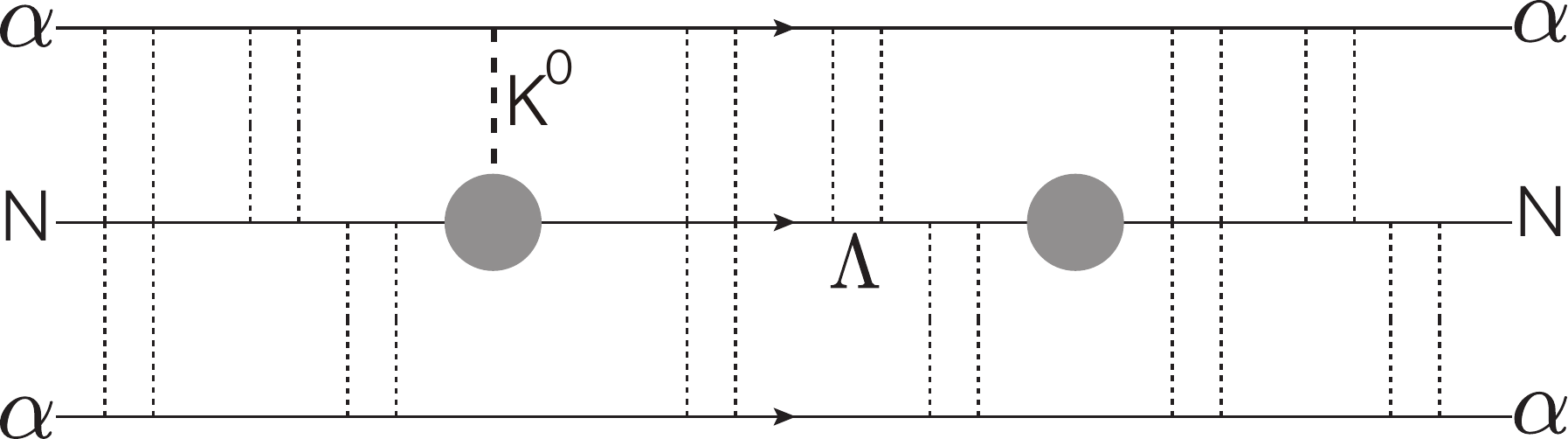}
\caption{The dynamical effect of the hypernuclear intermediate state.
The thick dashed line indicates the P-odd $|\Delta S|=1$
one-meson exchange interaction whereas the thin dashed lines denote
the interaction of the $\alpha$-$\alpha$ and $\alpha$-$N$ subsystems.
The gray blobs are the vertices of the
$|\Delta S|=1$ weak interactions.}
\label{fig:dy_hyperon}
\end{center}
\end{figure}

\begin{figure}[]
\begin{center}
\includegraphics[scale=0.45]{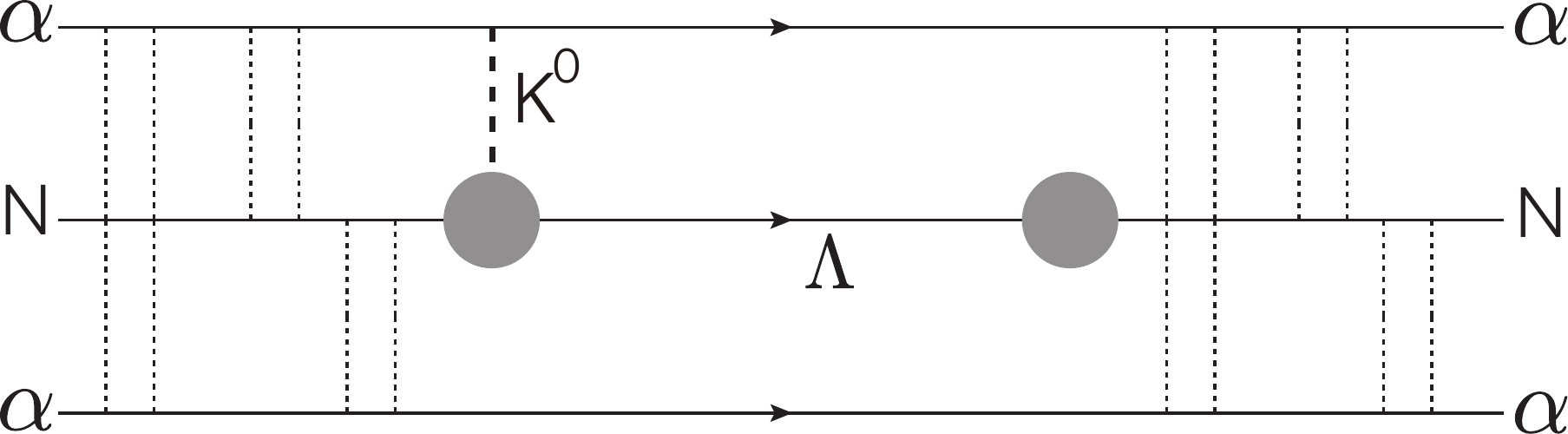}
\caption{The same as Fig. \ref{fig:dy_hyperon}, but without coupled channels.}
\label{fig:wo_dy_hyperon}
\end{center}
\end{figure}

To obtain the nuclear wave function of $^9$Be, we solve the nonrelativistic Schr$\ddot{\mathrm{o}}$dinger equation
\begin{align}\label{}
(H-E)\Psi_{JM_z} &=0,
\end{align}
where $J=m_z=\frac{3}{2}$.
Here, we use the Gaussian expansion method \cite{Hiyama:2003cu} to treat this problem.
In this framework, the wave function of  $^9$Be is given by
\begin{align}\label{}
\Psi_{JM_z, S}^{(c)}(\mathbf{r})
&= \mathcal{A}
\Bigg\{ \left[
\left[ \phi_{nlm}^{(c)}(\mathbf{r}_c) \varphi_{NLM}^{(c)}(\mathbf{R}_c) \right]_{\lambda}
\chi_{\frac{1}{2}} \right]_{J M_z}
\eta_{S}
\Bigg\},
\end{align}
where $\mathcal{A}$ is the antisymmetrization operator, $\chi$ and $\eta$ denote the spin and strangeness wave functions, respectively.
Here, we are considering the $NN$-$N \Lambda$ channel coupling which is spanned by the $S=0$ and $S=1$ wave functions.
With this basis, we can take into account the dynamical effect of the interaction among the hyperon and the other nucleons in the intermediate states (see Fig. \ref{fig:dy_hyperon}) which was neglected in the previous work \cite{Yamanaka:2015ncb} (see Fig. \ref{fig:wo_dy_hyperon}).
The wave functions are given as a superposition
of Gaussian basis functions
\begin{align}\label{}
\phi_{nlm}^{(c)}(\mathbf{r}_c)
&= N_{nl} r^{l}_c e^{-(r_c/r_n)^2} Y_{lm}(\hat{\mathbf{r}}_c), \\
\varphi_{NLM}^{(c)}(\mathbf{R}_c)
&= N_{NL} R^{L}_c e^{-(R_c/R_N)^2} Y_{LM}(\hat{\mathbf{R}}_c),
\end{align}
with the normalization constants $N_{nl}$ and $N_{NL}$.
The Gaussian range parameters are given
in a geometric progression
\begin{align}\label{}
r_n &= r_1 a^{n-1}  \;\;\;\;\;\;\;\, (n=1,...\,,n_{max}), \\
R_N &= R_1 A^{N-1}  \;\;\;\;\;\; (N=1,...\,, N_{max}).
\end{align}


\section{Electric dipole moment}\label{sec:EDM}

\begin{figure}[]
\begin{center}
\includegraphics[scale=0.3]{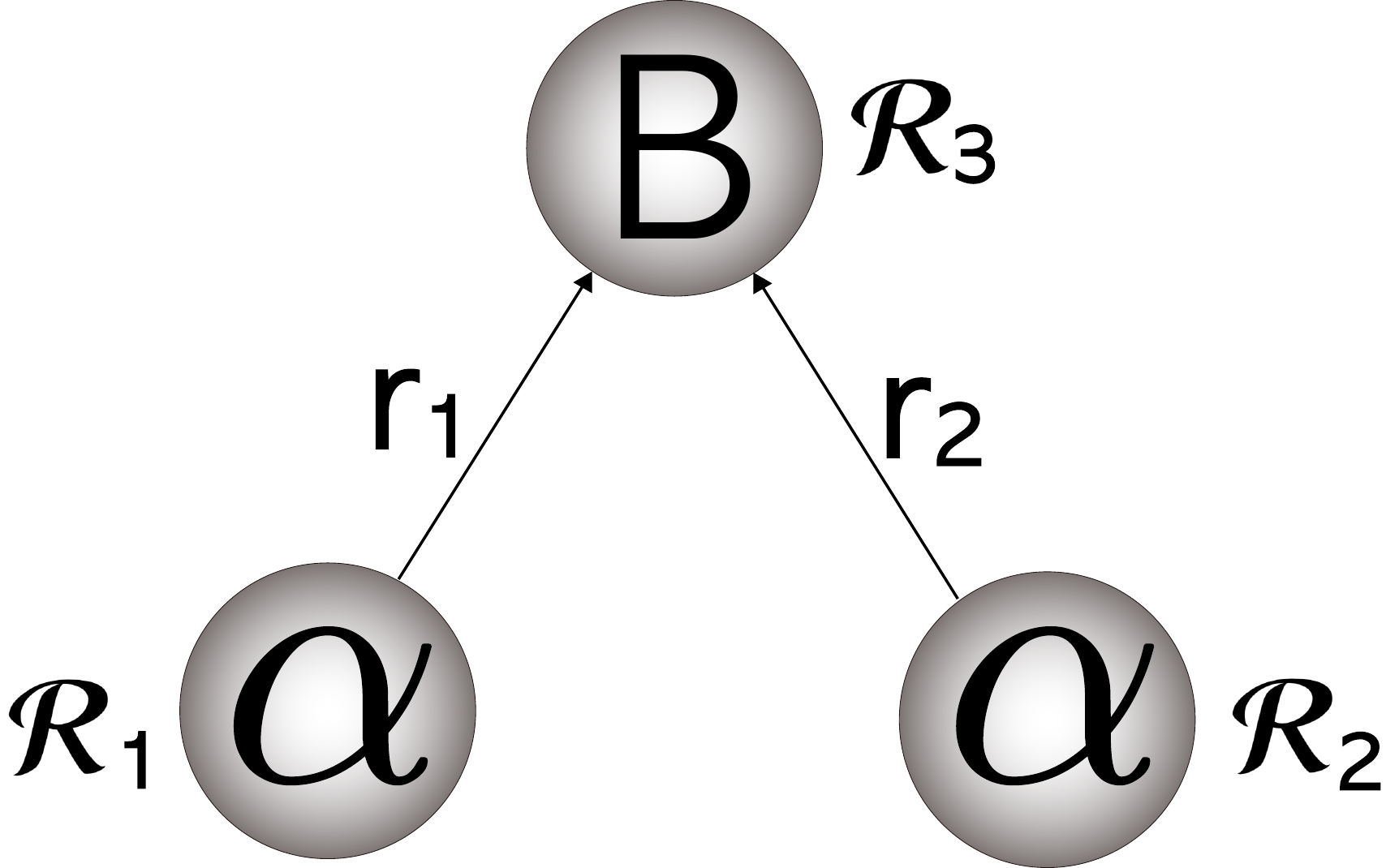}
\caption{The coordinates of the $\alpha \alpha B$ system where $B$ denotes $n$ or $\Lambda$.}
\label{fig:aaB}
\end{center}
\end{figure}

The electric-dipole operator
of $^9$Be in the $\alpha \alpha n$ cluster model is given by
\begin{align}\label{}
\mathcal{E} =
e \sum_{i} Q_i \mbox{\boldmath $\mathcal{R}$}_i  + (\mathbf{ C.M.})
&= -\frac{2}{9}e (\mathbf{r}_1 + \mathbf{r}_2),
\end{align}
where the center of mass vector is arbitrary and unphysical.
Here, the relative coordinates are defined as
\begin{align}\label{}
\mbox{\boldmath $\mathcal{R}$}_1 &= \mbox{\boldmath $\mathcal{R}$}_3 - \mathbf{r}_1, \\
\mbox{\boldmath $\mathcal{R}$}_2 &= \mbox{\boldmath $\mathcal{R}$}_3 - \mathbf{r}_2.
\end{align}
For illustration, see Fig. \ref{fig:aaB}.
Similarly, the dipole operator for
the $\alpha \alpha \Lambda$ system is given by
\begin{align}\label{}
\mathcal{D}=
e \sum_{i} Q_i \mbox{\boldmath $\mathcal{R}$}_i + (\mathbf{ C.M.})
&\approx -\frac{12}{46}e (\mathbf{r}_1 + \mathbf{r}_2).
\end{align}

\section{Results and discussion}\label{sec:results}
From our calculation, the contribution of the $|\Delta S|=1$
kaon exchange interaction to
the EDM of $^9$Be is obtained as
\begin{align}\label{d_Be9L1}
d_{^9\mathrm{Be}}
&
= 5.47 \times 10^{-32} \, e \, \mathrm{cm}
.
\end{align}
On the other hand, the $^9$Be EDM without considering the intermediate hypernuclear contribution is
\begin{align}\label{d_Be9L2}
d_{^9\mathrm{Be}}
&
= 5.8 \times 10^{-32} \, e \, \mathrm{cm}
,
\end{align}
calculated with the effective CP-odd $NN$ interaction, obtained by integrating out the intermediate $\Lambda$ \cite{Yamanaka:2015ncb}.
We see that the results are close.
This result strongly suggests that
the Pauli exclusion principle and the $YN$ interaction are not significant in the hypernuclear intermediate state for the nuclear EDM generated by the CKM matrix.

By separately calculating the first and the second contributions of Eq. (\ref{eq:long}), we have, respectively,

\begin{align}\label{}
d_{^9\mathrm{Be}}^{(\alpha \alpha n)}
&= 5.16 \times 10^{-32} \, e \, \mathrm{cm},\\
d_{^9\mathrm{Be}}^{(\alpha \alpha \Lambda)}
&= 0.31 \times 10^{-32} \, e \, \mathrm{cm}.
\end{align}
This result confirms our order estimation.
Equation (\ref{eq:long}) also suggests that we carefully inspect the EDM of deformed nuclei, since the enhancement due to the close energy levels of opposite parity states \cite{Lackenby:2018nco,Flambaum:2002mv,Flambaum:2018kuh,Dobaczewski:2005hz, Dobaczewski:2018nim,Engel:1999np, Engel:2003rz, Haxton:1983dq} is upset by taking the transition to the excited opposite parity states.

Another potential suppression mechanism to be inspected is the nonperturbative effect contributing to the matrix element in the numerator (this corresponds to the contribution which needs to be resumed in the calculation of the neutrinoless double-beta decay \cite{Cirigliano:2017tvr}).
Indeed the difference of structure between $^9$Be and $^9_{\Lambda}$Be may significantly damp it.
In our calculation, the $N \alpha$ distance for $^9\mathrm{Be}(3/2^-)$ is 3.6 fm and the distance of $\Lambda \alpha$ for $^9_{\Lambda}\mathrm{Be}(3/2^+)$ is 4.8 fm, so the suppression should be relevant.
However, the other hypernuclear continuum states have good overlaps with $^9\mathrm{Be}(3/2^-)$ and the transition with them gives the leading contribution to the EDM.
This is why Eq. (\ref{d_Be9L1}) is not significantly suppressed compared with Eq. (\ref{d_Be9L2}).
We note that the difference of energies between the continuum states and the bound states of $^9_{\Lambda}$Be is much smaller than the hyperon-nucleon mass difference.
The transition through excited states are therefore not suppressed by the denominator.
Overall, we may say that the EDM of $^9$Be receives a contribution from higher virtual states.

\begin{figure}[]
\begin{center}
\includegraphics[width=\linewidth]{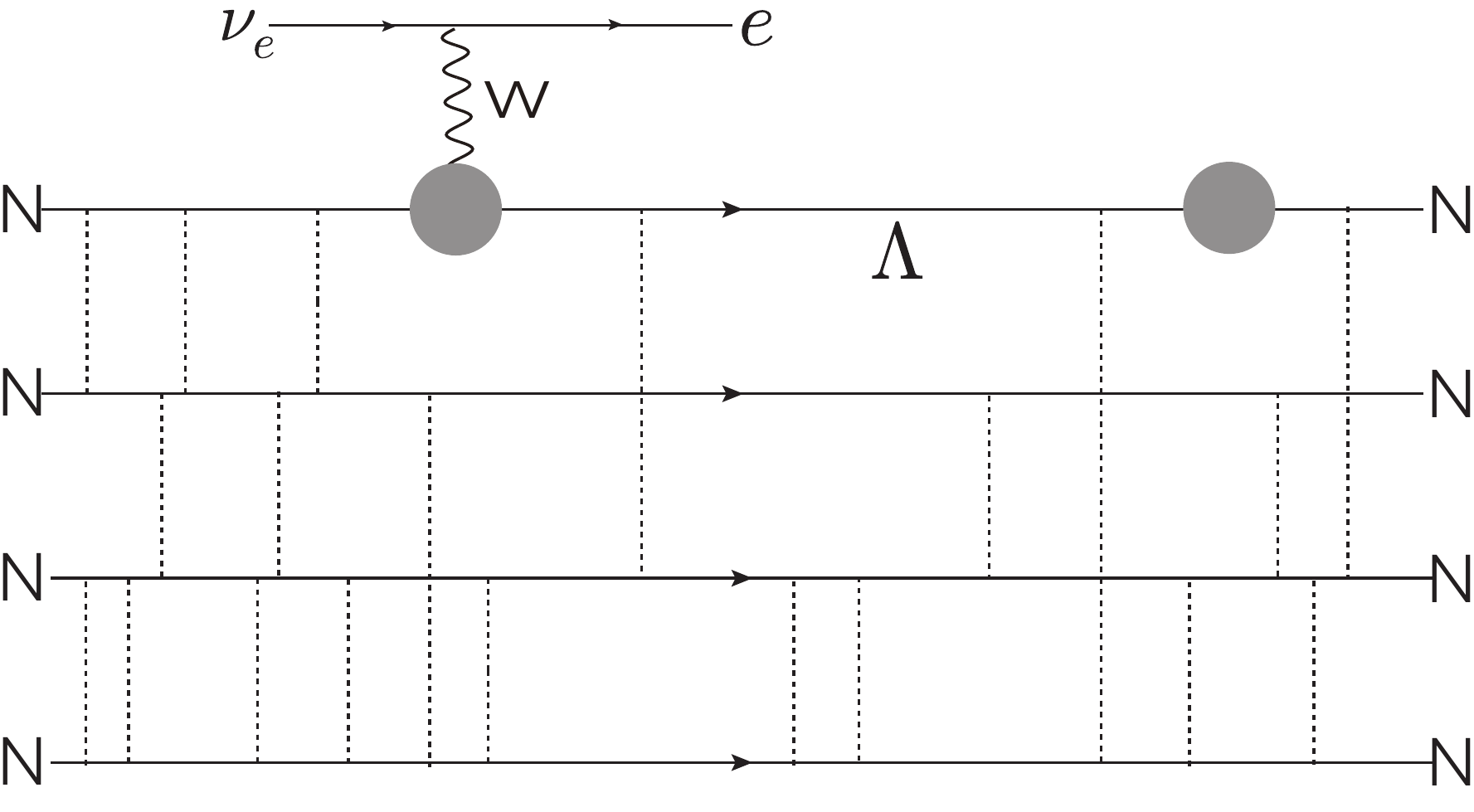}
\caption{
Diagrammatic representation of the nuclear beta decay with hypernuclear intermediate states in the standard model. The gray blobs denote the $|\Delta S|=1$ interaction.
The $\Lambda$ hyperon is interacting with other nucleons in the nucleus.
}
\label{fig:beta-decay}
\end{center}
\end{figure}

Let us also calculate
the matrix element of the one-body $\Lambda$-$n$ transition
to compare with the case without consideration of the nucleus-hypernucleus mixing.
By taking the ratio between the cases with and without channel coupling, we obtain
\begin{align}\label{}
\frac{ 2 \langle \Psi | n \Lambda^{\dagger}| \Psi \rangle}
{ \mathrm{Re}(a_{n \Lambda})
\langle ^9\mathrm{Be}| n^{\dagger} \Lambda
| \alpha \alpha \Lambda \rangle
\frac{1}{m_n - m_{\Lambda}}
\langle \alpha \alpha \Lambda| n \Lambda^{\dagger}
| ^9\mathrm{Be}\rangle
}
=0.99.
\end{align}
We see that the ratio is very close to one.
\begin{figure}[]
\begin{center}
\includegraphics[scale=0.7]{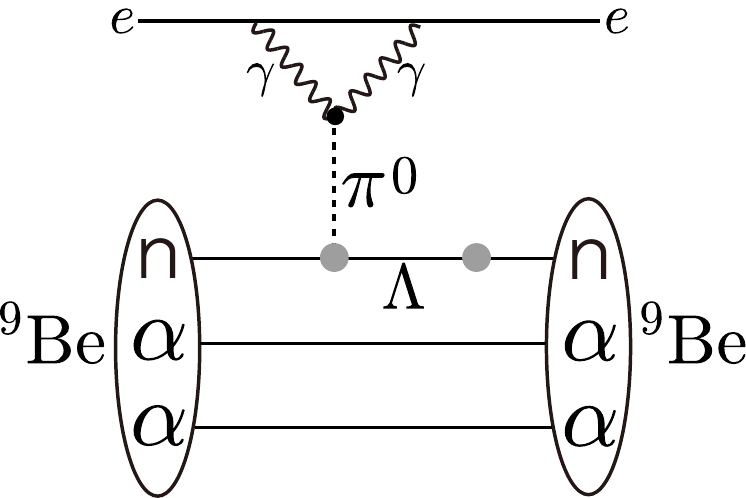}
\caption{
CP-odd electron-nucleon interaction with $\pi^0$ exchange, contributing to the atomic EDM.
}\label{fig:cp_odd}
\end{center}
\end{figure}
This means that the distributions of the $\alpha \alpha n$ and
the $\alpha \alpha \Lambda$ states closely resemble each other.
The evaluation of this hyperon-nucleon transition matrix element is also important to quantify the nuclear effect in the CP-odd electron-nucleon interaction which is one of the leading contribution to the EDM of atoms.
At the leading order, it is generated by the $|\Delta S|=1$ meson-baryon interaction and the hyperon-nucleon transition which fulfills the Jarlskog combination \cite{Pospelov:2013sca}, but the meson generated by the meson-baryon interaction is connected to the outer electrons, so that the two $|\Delta S|=1$ interactions act as a one-body process at the nuclear level (see Fig. \ref{fig:cp_odd}).
The $^9$Be nucleus cannot be used in atomic EDM experiments, since it is too light to enhance the CP violation.
However, our result suggests that the CP-odd electron-nucleon interaction generated by the CKM matrix is not suppressed in heavier nuclei, since the Pauli exclusion principle does not have significant effect in the hypernuclear intermediate state.
The same remark applies for the $T$-odd angular correlations of the nuclear beta decay which also has hypernuclear intermediate states (see Fig. \ref{fig:beta-decay}).

The theoretical uncertainties of this calculation are the followings:
they are due to
(i) the renormalization-group evolution of the $|\Delta S|$ = 1 four-quark interactions,
(ii) the hadronic matrix elements, and
(iii) the systematics at the nuclear level.
Let us see them in detail.
Regarding (i), the most important error is due to the nonperturbative effect near the hadronic scale.
In Ref. \cite{Yamanaka:2015ncb}, it was deduced that the results may change by a factor of two by varying the renormalization scale, but the order of magnitude does not.
The electroweak contribution in this context is not important.
Concerning (ii), we have the calculation of the matrix elements of the hyperon-nucleon transition and that of $|\Delta S|$=1 meson-baryon interactions.
The former has been obtained from the fit of the nonleptonic hyperon decay \cite{Hiyama:2004wd} so the error bar should not be large.
However, the evaluation of the meson-baryon interaction required the QCD calculation of the matrix elements of the subleading four-quark operators $Q_5^q$ and $Q_6^q$ (see Fig. \ref{fig:w_ex} (b)).
In this work, we used the factorization which has a large theoretical uncertainty (in Ref. \cite{Yamanaka:2016fjj}, it is estimated to be of $O(60\%)$).
Moreover, there is an additional systematic error due to the pion-nucleon sigma term (see Sec. \ref{sec:quark}).
In view of this, the hadron level systematics might affect the order of the magnitude of our results.

The remaining uncertainty is the nuclear-level uncertainty.
It comprises the systematics of the CP-even interactions of the $\alpha$-cluster model,
the folding of the $|\Delta S|=1$ two-body interaction, the contribution from the intrinsic nucleon EDM, and the effect beyond the $\alpha$-cluster model.
The first one should not be important because low-lying energy levels of $^9$Be and $^9_{\Lambda}$Be are well described within the model we adopted while the nuclear EDM is sensitive to the long distance and low energy physics.
The second one has never been quantified, but we expect it to be subleading since the $K$ meson is relatively light.
The effect of the intrinsic nucleon EDM has been estimated in a previous work to be $O(10^{-32})e$ cm \cite{Seng:2014lea}, so it may interfere with our result (\ref{d_Be9L1}).
Its quantification is very difficult at the present stage due to the poor knowledge of the low energy constants of chiral perturbation theory.
However, the most important systematics is probably the effect of the transition of the nucleon inside the $\alpha$-cluster to the hyperon, since the pion exchange, which was the leading contribution in the case without consideration of the nucleus-hypernucleus mixing, becomes relevant.
We know from the study of the EDM of $^{13}$C that, the large energy required in the transition between states suppresses the amplitude of the process.
From this point-of-view, the destruction of the $\alpha$-cluster may be important, since the binding energy of the $^4$He nucleus is large (28.3 MeV).

\section{Summary}\label{sec:summary}
In this paper we have studied the contributions of the $|\Delta S|=1$
$K$ meson exchange process generated by the CKM matrix to the EDM of the $^9$Be by considering the $\alpha n$-$\alpha \Lambda$ channel coupling within the GEM.
We have found that the result of the $^9$Be EDM obtained by considering the hypernuclear intermediate states does not differ by much from the result with the $\Lambda$ hyperon integrated out.
We conclude that the Pauli exclusion principle in the hypernuclear intermediate states does not have a significant effect in the nuclear EDM.
This is probably due to the large mass difference between the hyperon and the nucleon compared with the typical binding effect in nuclei or hypernuclei, which brings high virtuality in the intermediate hypernuclear states.

The important uncertainty is the contribution beyond the $\alpha$-cluster model.
The destruction of the $\alpha$-cluster will lead to loss of 28.3 MeV which will partially cancel the virtuality due to the intermediate hyperon.
To evaluate this effect, the EDM of $^9$Be with a nine-body {\it ab initio} calculation is required.
This is left for future work.

The change of the EDM generated by the $|\Delta S|=1$ interactions has an important impact not only to the CKM contribution, but also to flavor-violating models beyond the standard model \cite{Ellis:1981ts,
Cheng:1987rs,
Hagelin:1992tc,
Gabbiani:1996hi,
Buchalla:1996fp,
Fritzsch:1999ee,
DAmbrosio:2002vsn,
Buras:2003td,
Agashe:2004cp,
Antonelli:2009ws,
Isidori:2010kg,
Buras:2010zm,Charles:2015gya,
Kitahara:2016nld,Kitahara:2016otd,Cirigliano:2016yhc,Endo:2017ums,Smith:2017dtz,Chang:2017wpl,Agrawal:2017evu,Aebischer:2018csl,Dekens:2018bci,
Gisbert:2017vvj, Haba:2018byj, Chen:2018ytc, Chen:2018vog, Bailey:2018feb,Haba:2018rzf}.
It is then important to further quantify the systematics due to the nucleus-hypernucleus mixing.

\section*{Acknowledgements}
This work was supported by RIKEN Junior Research Associate (JRA) scholarship and by the Grant-in-Aid for Scientific Research (Grants No.16H03995) from the Japan Society For the Promotion of Science. N.Y. was supported by the JSPS Postdoctoral Fellowships for Research Abroad.

\end{document}